 \documentclass[conference]{IEEEtran}
 \pdfoutput=1 

\usepackage{cite}
\usepackage{multirow}
\usepackage{graphicx}
\usepackage{amssymb,amsmath,times,wasysym}
\usepackage{pb-diagram}
\usepackage{psfrag,balance}
\usepackage{srcltx,verbatim} 
\newcounter{mycounter}
\usepackage{fixltx2e}
\usepackage{multicol}
\usepackage{color}
\usepackage{algorithmic}
\usepackage{algorithm}
\usepackage[caption=false]{subfig}

\newcounter{appendx}

\renewcommand{\exp}[1]{\mathrm{e}^{#1}}

\renewcommand{\log}[2][]{\mathrm{log}_{#1}\left(#2\right)}

\newcommand{\Q}[1]{\mathcal{Q}\left(#1\right)}

\newcommand{\cdf}[2][]{F_{#1}\!\left({#2}\right)}

\newcommand{\E}[2][]{\mathbb{E}_{#1}\!\left[{#2}\right]}

\newcommand{\Var}[1]{\mathbb{V}\mathrm{ar}\!\left[{#1}\right ]}

\newcommand{\e}[1]{\text{exp}\!\left({#1}\right)}

\newcommand{\GammaFn}[1]{\Gamma\!\left({#1}\right)}

\title{{Performance Analysis of Intelligent Reflective Surfaces for Wireless Communication} \vspace{-5mm}}

\author{\IEEEauthorblockN{Dhanushka Kudathanthirige, Dulaj Gunasinghe and Gayan Amarasuriya } 
	\IEEEauthorblockA{Department of Electrical and Computer    Engineering, Southern Illinois University, Carbondale, IL, USA 62901\\Email: \{dhanushka.kudathanthirige,dulaj.gunasinghe,gayan.baduge\}@siu.edu \vspace{-7mm}}
\thanks{This paper has been accepted to present at IEEE International Communication Conference (ICC) 2020, Dublin, Ireland.}

}

\DeclareMathAlphabet\mathbfcal{OMS}{cmsy}{b}{n}
\IEEEoverridecommandlockouts
\begin{document}
\bstctlcite{IEEEexample:BSTcontrol}
\vspace{-5mm}
\maketitle
 
\begin{abstract}

	A statistical characterization of the fundamental performance bounds of an intelligent reflective surface (IRS) intended for aiding wireless communications is presented. To this end, the outage probability, average symbol error probability and   achievable rate bounds are derived in closed-form. By virtue of an asymptotic analysis in  high signal-to-noise ratio (SNR) regime, the achievable diversity order is derived. Thereby, we show that   a diversity gain   in the order of the number of passive reflective elements embedded  within the IRS  can be achieved with only controllable phase adjustments. Thus, IRS has a great  potential of boosting the wireless performance by intelligently controlling the  propagation channels  without employing additional active radio frequency chains.

\end{abstract}


\vspace{-1mm}
\section{Introduction}\label{sec:introduction}

Over the past five generations of wireless standards,  performance of the  transmitter and receiver has been optimized 
to mitigate various transmission impairments of propagation channels, which are generally assumed to be uncontrollable in  the wireless system designer's  perspective. However, owing to the recent research advancements of meta-materials and meta-surfaces,  a novel concept of coating physical objects such as building walls and windows with intelligent reflective surfaces (IRSs) with reconfigurable  reflective  properties has  been envisioned \cite{Liaskos2018,Su2017,Yang2016,Renzo2019}. The ultimate goal of IRS is to enable a smart wireless propagation  environment by controlling the reflective properties of the  underlying channels \cite{Renzo2019}.

   An IRS comprises of  a very large number of passive reflective elements, which are capable of  reconfiguring  properties of electromagnetic (EM) waves impinging upon them.    On one hand, reflected EM waves can be added constructively at a  desired  receiver by intelligently controlling  phase-shifts at each reflective element to boost the signal-to-noise ratio (SNR) and coverage. On the other hand, a   reflected  signal can be made to    add  destructively and thereby to mitigate co-channel interference towards an undesired direction.
  Moreover, IRS   facilitates full-duplex reflections, and hence, large blockages between a pair of transmitter-receiver can be circumvented through smart reflections without trading-off additional time, frequency or power resources.       
  Since an IRS does not generate new EM waves, costly transmit radio-frequency (RF) chains/amplifiers in relays can be eliminated and thereby improving the energy efficiency. 
  Thus, the concept of IRS presents  a paradigm shift in wireless communication research.

   \noindent\textbf{Prior related research:} 
     Due to recent breakthroughs  in physics and related fields \cite{Lee2012,Sekitani2009,Cui2015}, the designs of software-controllable  IRS have been shown to be feasible, and     core technical aspects  are currently being developed \cite{Liaskos2015,Liaskos2018}. The prototypes of meta-surfaces and meta-tiles with artificial thin film of EM  materials, which can be used to coat objects within a   smart wireless environment, have already been developed \cite{Lee2012,Sekitani2009}. 
   In  \cite{Wu2019}, precoder optimization techniques for a multi-antenna transmitter in the presence of an IRS are investigated to maximize the received SNR. In  \cite{Basar2019}, basic ray tracing techniques are adopted to model multipath propagation through an IRS, and thereby, it discusses techniques for controlling the reflections via controllable phase-shifts at passive elements embedded within an IRS. Moreover,   in \cite{Emil2019}, transmit power scaling laws pertaining to IRS are derived to alleviate misconceptions about the performance comparisons between the IRS and massive multiple-input multiple-output (MIMO) systems.  
     In \cite{Han2019},    an optimal phase shift design is proposed for IRS 
   based on maximizing an  upper bound of the average spectral efficiency. 
   In \cite{Chen2019}, techniques for boosting the physical layer security through smart propagation enabled by   an IRS are investigated.

    \noindent\textbf{Motivation and our contribution:} 
      The key idea of an IRS is  to enable a programmable control over the wireless propagation channels.      This necessitates  innovations of  radically  novel techniques for modeling, designing and analyzing wireless systems as the resulting  smart propagation channels can now be able to interact with    EM waves impinging upon them in a software-controlled manner. Although  several important attempts  have recently been made \cite{Wu2019,Basar2019,Emil2019,Han2019,Chen2019}, the fundamental research on IRS in wireless communication's perspective is still at an embryonic stage. To this end, our work presents a performance analysis framework for deriving the fundamental bounds pertaining to an IRS intended for aiding the end-to-end communication between a single-antenna source ($S$) and a destination ($D$).  Thus, tight bounds/approximations for the outage probability, average symbol error rate (SER), and average achievable rate are derived in closed-form. The accuracy of our analysis is validated through a rigorous set of Monte-Carlo simulations.  We obtain useful design insights about the achievable diversity order via an asymptotic analysis in the high SNR regime. We reveal that the achievable diversity order is equal to the number of passive reflective elements ($M$), and it is achieved without using any active RF chains at the IRS. Thus, by virtue of smart passive reflections, an $M$-fold diversity gain can be achieved with respect to a single-input single-output (SISO) channel. Through our analysis, we verify that an IRS has a true potential of boosting the reliability of end-to-end wireless communication with only passive controllable phase-shifts.

 \noindent \textbf{Notation:} $\mathbf x^{T}$ denotes the transpose of  $\mathbf x$.
 $\E[]{X}$ and $\Var{X}$ represent the expectation and variance of  a random variable (RV) $X$, respectively.
  $X\sim\mathcal {CN}\left(\mu_X,  \sigma_X^{2}  \right) $ denotes that $X$ is complex Gaussian distributed with $\mu_X$ mean and $\sigma_X^{2}$ variance. 

 \section{System, channel and signal models}\label{sec:system_model}
 
 \begin{figure}[!t]\centering \vspace{0mm}
 	\def\svgwidth{200pt} 
 	\fontsize{8}{4}\selectfont 
 	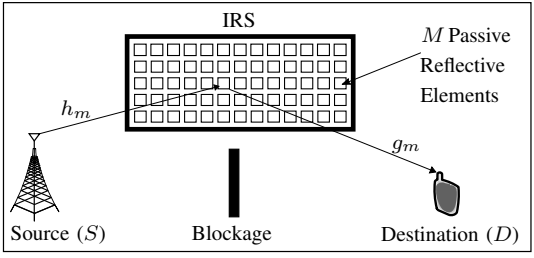 \vspace{0mm}
 	\caption{An IRS-assisted  wireless communication set-up.}\vspace{-6mm} \label{fig:system_model}
 \end{figure}


\subsection{System model}
We consider an end-to-end wireless communication set-up in which a single-antenna source ($S$) communicates with a single-antenna destination ($D$) through    an IRS having  $M$-passive reflective elements (see Fig. \ref{fig:system_model}). Phase-shifts of waves impinging upon the IRS are assumed to be  controlled perfectly to implement   coherent/constructive  signal combining at $D$, while the corresponding  amplitudes are attenuated by a factor defined as per reflective coefficient. 
 The direct channel between $S$ and $D$ is assumed to be unavailable due to severe blockage effects. The channel  between  $S$ and the $m$th reflective element is denoted by $h_m$, while the channel between the  $m$th reflective element and $D$ is  given by  $g_m$. The channel envelopes are assumed to be    independent Rayleigh distributed, and hence, $h_m$ and $g_m$   are modeled as
 
\vspace{-3mm}
\begin{small} 
 \begin{eqnarray}\label{eqn:channel_model_h}
 h_m = \sqrt{\zeta_{h_m}}\tilde{h}_m\qquad \text{and}\qquad 
 g_m = \sqrt{\zeta_{g_m}}\tilde{g}_m,
 \end{eqnarray}
\end{small}
\vspace{-4mm}
 
\noindent where $\tilde{h}_m$ and $\tilde{g}_m$ follow   complex Gaussian distribution with zero mean and  unit variance; $\tilde{h}_m\sim \mathcal{CN}(0,1)$ and $\tilde{g}_m \sim \mathcal {CN}(0,1)$. In (\ref{eqn:channel_model_h}), $\zeta_{h_m}$ and $\zeta_{g_m}$  capture the path-losses of the corresponding channels. 

 \subsection{Signal model}
 The signal transmitted by $S$ is reflected by the IRS towards $D$. 
The received signal at $D$ through $M$ reflective elements can be written as 

\vspace{-3mm}
\begin{small}
	 \begin{eqnarray}\label{eqn:received_signal}
	y =\sqrt{P} \displaystyle \sum_{m=1}^{M} g_m \eta_m \exp{j\theta_m} h_m x+ n,
	\end{eqnarray}  
\end{small}
\vspace{-4mm}
 
\noindent
where $x$ is the transmitted signal by $S$ satisfying $\E{|x|^2}=1$, while  $P$ is the transmit power at $S$.  Moreover, $n$ is an additive white Gaussian noise (AWGN) at D with zero mean and variance $\sigma_N^2$ such that $n\sim \mathcal {CN}(0, \sigma_N^2)$. In (\ref{eqn:received_signal}), $\eta_m$ and $\theta_m$ represent the reflection coefficient and the phase-shift introduced by the $m$th reflective component of the IRS, respectively. 
Next, (\ref{eqn:received_signal}) can be  alternatively written as \cite{Wu2019}

\vspace{-5mm}
\begin{small}
  \begin{eqnarray}\label{eqn:rx_signal_matrix}
 y =\sqrt{P} \mathbf{g}^T \mathbf{\Theta} \mathbf{h} x   + n,
 \end{eqnarray} 
 \end{small}
\vspace{-5mm} 

\noindent where $\mathbf{h}$ = [$h_1$, $\cdots$, $h_m$, $\cdots$, $h_M$]$^{T}$ and $\mathbf g$ = [$g_1$, $\cdots$,  $g_m$, $\cdots$, $g_M$]$^{T}$ capture the corresponding channel vectors, while  $\mathbf{\Theta}=\text{diag}([\eta_1 \exp{j\theta_1}, \cdots, \eta_m \exp{j\theta_m},  \cdots, \eta_M \exp{j\theta_M}])$ is a  diagonal matrix, which captures  the reflection properties of $M$ reflective elements of the IRS.

By using (\ref{eqn:rx_signal_matrix}), the SNR is derived  as

\vspace{-3mm}
\begin{small}
  \begin{eqnarray}\label{eqn:SNR_basic}
 {\Gamma} = { \bar{\gamma}| \mathbf{g}^T \mathbf{\Theta} \mathbf{h} | ^2 } ,
 \end{eqnarray}
\end{small}
 \vspace{-5mm} 
  
\noindent where $\bar{\gamma} = P/\sigma_N^2$ is the average transmit SNR. The channels    $h_m$ and $g_m$ in (\ref{eqn:channel_model_h}) can be rewritten as 

\vspace{-3mm}
\begin{small}
 \begin{eqnarray}\label{eqn:channel_model_h_ray}
 h_m = \nu_m \e{j \phi_{m}} \qquad\text{and}\qquad 
 g_m = \rho_m \e{j \omega_{m}},
 \end{eqnarray}
\end{small}
\vspace{-5mm} 

\noindent where $\nu_m = |h_m|$ and $\rho_m = |g_m|$ are the channel amplitudes with  Rayleigh distribution,  while $\phi_{m}$ and $\omega_{m}$ are the corresponding channel phases, which are  uniformly distributed  between $\left(\right.\!\!-\pi, \pi\left.\right]$. By substituting  (\ref{eqn:channel_model_h_ray}) into (\ref{eqn:SNR_basic}), the SNR in (\ref{eqn:SNR_basic}) can be expanded as 

\vspace{-4mm}
\begin{small}
 \begin{eqnarray}\label{eqn:SNR}
{\Gamma} = { \bar{\gamma}\left| \displaystyle \sum_{m=1}^{M} \rho_m \nu_m \eta_m \e{j[ \phi_{m}+\omega_{m}+\theta_m]} \right| ^2 } .
\end{eqnarray}
\end{small}
\vspace{-4mm} 
  
\noindent The $M$ terms inside the summation of (\ref{eqn:SNR}) must be constructively added to maximize the received SNR. This can be accomplished by intelligently controlling the reflective properties ($\theta_m$) of each element within the IRS.  More specifically, $\Gamma$ in (\ref{eqn:SNR}) can be maximized by co-phasing each term in its summation. Thus, the optimal choice   of $\theta_m$ is given by $ \underset {-\pi <  \theta_m \leq  \pi  }{\text{arg}\!\max}  \;\{\Gamma\} =  \theta^*_m = -(\phi_{m}+\omega_{m})$ \cite{Wu2019}, and thereby, the optimal received SNR at $D$  can be written as

\vspace{-4mm}
\begin{small}
 \begin{eqnarray}\label{eqn:SNR_Optimal}
 \Gamma^* =   { \bar{\gamma}\left| \displaystyle \sum_{m=1}^{M} \rho_m \nu_m  \eta_m\right| ^2 } .
\end{eqnarray} 
\end{small}
\vspace{-4mm}

\section{Performance Analysis}\label{sec:performance_analysis}

\begin{figure}[!t]\centering\vspace{-3mm}
	\includegraphics[width=0.4 \textwidth]{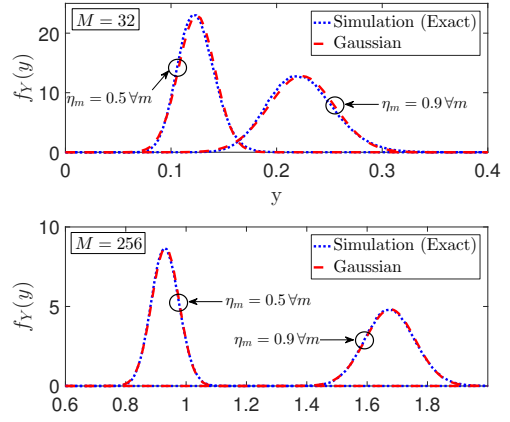}\vspace{-5mm}
	\caption{The exact PDF of $Y=\sum_{m=1}^{M} \rho_m \nu_m  \eta_m$ via Monte-Carlo simulations and  its analytical approximation via CLT.}
	\label{fig:PDF}\vspace{-5mm}
\end{figure}

\subsection{Statistical characterization of the optimal received SNR}

We notice that $\rho_m$ and $\nu_m$ are independently Rayleigh distributed RVs   for $m\in \{1,\cdots, M\}$. Thus, according to the central limit theorem (CLT), $Y=\sum_{m=1}^{M} \rho_m \nu_m  \eta_m$ coverages to a Gaussian  distribution for a sufficiently large number of  passive elements in the  IRS \cite{Papoulis2002}. 
Thereby, the distribution of $\Gamma^*$ in (\ref{eqn:SNR_Optimal}) can be tightly  approximated by a non-central chi-squared distribution with a single degree-of-freedom \cite{Papoulis2002}; $\Gamma^* \approx \gamma \sim \mathcal X^2_1(\nu)$, where $\nu$ is the non-centrality parameter. 
The accuracy of this approximation is verified via the probability density function (PDF) curves  in Fig. \ref{fig:PDF} for different $M$ and $\eta_m$. 
Consequently, the cumulative distribution function (CDF) of $\gamma$ can be written as (see Appendix \ref{app:AppendixAA})

\vspace{-4mm}
\begin{small}
\begin{eqnarray}\label{eqn:Optimal_SNR_CDF}
F_{\gamma}(z) =
1 - \psi\Q{\left({\sqrt{ z/\bar{\gamma} }-\mu_Y}\right)\Big/ {\sigma_Y}},
\end{eqnarray} 
\end{small}
\vspace{-4mm} 

\noindent   where $z\geq 0$, and  $\Q{\cdot}$ denotes the Gaussian $\mathcal Q$ function,   which is defined as $\Q{x}=\int_{x}^{\infty}\exp{-t^2/2}\big/{\sqrt{2\pi}}dt$ \cite{Proakis2007}. 
%
%
\noindent In (\ref{eqn:Optimal_SNR_CDF}), $\mu_Y$, $\sigma_Y$, and $\psi$  are given by

\vspace{-4mm}
\begin{small}
\begin{subequations}
	\begin{eqnarray}
		\!\!\!\!\!\!\!\!\!\! \mu_Y &=& \sum_{m=1}^M\pi\lambda_m/2 \label{eqn:mean}  \quad \text{and}\quad
	 \sigma^2_Y = \sum_{m=1}^M\lambda^2_m \left[16-{\pi}^2\right]/4,\;\label{eqn:varaiance1} \\
		\!\!\!\!\!\!	\!\!\!\!\!\! \psi&=&1/\left[\Q{-\sqrt{\kappa}}\right], \label{eqn:varaiance}
	\end{eqnarray} 
\end{subequations}
\end{small}
\vspace{-4mm}

\noindent where $\lambda^2_m=\zeta_{h_m}\zeta_{g_m}\eta_m^2/4$, $\kappa=\mu^2_Y/\sigma^2_Y$ and $\GammaFn{t} = \int_{0}^{\infty} x^{t-1}\exp{-x}dx$ is the Gamma function \cite[Eqn. (8.310.1)]{Gradshteyn2007}.

\subsection{Outage Probability}\label{sec:outage}
The SNR outage is defined as the probability that the instantaneous SNR $(\gamma)$ falls bellow a threshold SNR   $(\gamma_{th})$. By using   (\ref{eqn:Optimal_SNR_CDF}), a tight approximation for the outage probability in    moderately large $M$ regime  can be  obtained as 

\vspace{-4mm}
\begin{small}
\begin{eqnarray}\label{eqn:P_out}
P_{out}  = \Pr(\Gamma^*\leq \gamma_{th} ) \approx \cdf[\gamma]{\gamma_{th}},
\end{eqnarray}
\end{small}
\vspace{-4mm} 

\noindent  where $\cdf[\gamma]{\cdot}$ is defined in (\ref{eqn:Optimal_SNR_CDF}).

 \noindent \textit{\textbf {Remark 1:}} The PDF of $\Gamma^*$ in (\ref{eqn:SNR_Optimal}) for $M=1$ is given by $f_{\Gamma^*}(z)=(2\bar{\gamma}\lambda^2_1)^{-1}K_0(\sqrt{z/(\bar{\gamma}\lambda^2_1}))$ \cite{Salo2006},  and the exact outage probability can be derived by using   \cite[Eqn. (5.56.2)]{Gradshteyn2007} as 
\begin{eqnarray}
P_{out}=\cdf[\Gamma^*]{\gamma_{th}}=1-\sqrt{{\gamma_{th}}/{\bar{\gamma}\lambda^2_1}} K_{1}(\sqrt{{\gamma_{th}}/{\bar{\gamma}\lambda^2_1}}),
\end{eqnarray}
where $ K_{v}\left(\cdot\right)$ is the $v$th order modified Bessel function of the second kind \cite[Eqn. 8.407.1]{Gradshteyn2007}.

\vspace{-4mm}
\subsection{Average achievable rate}
The average achievable  rate can be defined as 

\vspace{-4mm}
\begin{small}
\begin{eqnarray}\label{eqn:Optimal_SNR_Rate}
\mathcal R = \E{\log[2]{1+\Gamma^*}} \approx \E{\log[2]{1+\gamma}}.
\end{eqnarray}  
\end{small}
\vspace{-4mm} 

\noindent The exact derivation of the expectation in (\ref{eqn:Optimal_SNR_Rate}) appears mathematically intractable.  Thus, we resort to upper and lower bounds by invoking Jensen's inequality as $\mathcal R_{lb} \apprle	\mathcal R\apprle \mathcal R_{ub} $ \cite{Zhang2014},
%
%
where $\mathcal R_{lb}$ and $\mathcal R_{ub} $  are defined as  

\vspace{-4mm}
\begin{small}
\begin{eqnarray}
 	\mathcal R_{lb}&=&{\log[2]{1+\left[\E{{1}\big/{\gamma}}\right]^{-1}}},\label{eqn:lwerbound}\\
 	\mathcal R_{ub} &=&\log[2]{1+\E{\gamma}}\label{eqn:upperbound}.
\end{eqnarray}
\end{small}
\vspace{-5mm} 

\noindent  The upper bound in (\ref{eqn:upperbound}) can be derived as (see Appendix \ref{app:AppendixBB})

\vspace{-4mm}
\begin{small}
	 \begin{eqnarray}\label{eqn:Optimal_SNR_Rate_ub}
	 \mathcal R_{ub} &=&\log[2]{1+ \left(\bar{\gamma}\sum_{m=1}^M\lambda^2_{m}\right)\left[\frac{(16-\pi^2)(1+\kappa)}{4}\right] }.
	 \end{eqnarray} 
\end{small}
\vspace{-4mm}

\noindent Next, a tight approximation for the lower bound in (\ref{eqn:lwerbound}) can be derived  as (see Appendix \ref{app:AppendixBB})

\vspace{-5mm}
\begin{small}
	\begin{eqnarray}\label{eqn:lowerbound_ana}
	\!\!\!\!\!\!\!\mathcal R_{lb} \approx\log[2]{\!1\!+\left(\bar{\gamma}\sum_{m=1}^M\lambda^2_{m}\right)\left[\frac{(16-\pi^2)(\kappa+1)^3}{4(\kappa^2+6\kappa+3)}\right]  }.
	\end{eqnarray} 
\end{small}
\vspace{-4mm}

%


\subsection{Average symbol error rate (SER)}
The average SER is defined as the expectation of  conditional error probability $(P_{e|\Gamma^*})$ over the distribution of $\Gamma^*$ \cite{Proakis2007}. 
For wide range of modulation schemes,   $P_{e|\gamma}$  is given by   $P_{e|\Gamma^*}=\alpha\Q{\sqrt{\beta\Gamma^*}}$, where $\alpha$ and $\beta$ are modulation   dependent fixed parameters \cite{Proakis2007}. In this context,   the average SER can be derived as 
$\bar{P}_e=\E{\alpha\Q{\sqrt{\beta\Gamma^*}}}$. By using (\ref{eqn:Optimal_SNR_CDF}) and by evaluating the expectation, a tight approximation for $\bar{P}_e$  can be derived as  (see Appendix \ref{app:AppendixCC})

\vspace{-4mm}
\begin{small}
	 \begin{eqnarray}\label{eqn:Pe}
	 \bar{P}_e &\approx& \E{\alpha\Q{\sqrt{\beta\gamma}}}\nonumber \\ &=&\frac{\alpha{\psi}\mathrm{exp}(-\mu_Y^2/2\sigma_Y^2)}{\sqrt{2}{\pi}\sigma_Y}\int_{0}^{\pi/2}\frac{\mathrm{exp}\left(\mu_Y^2\big/\left(\frac{2\beta\bar{\gamma}\sigma_Y^4}{\sin^2\theta}+2\sigma_Y^2\right)\right)}{\sqrt{\frac{\beta\bar{\gamma}}{2\sin^2\theta}+\frac{1}{2\sigma_Y^2}}}\nonumber\\
	 &&\quad \quad \quad \times \Q{{\mu_Y}\big/{\sqrt{\frac{\beta\bar{\gamma}\sigma_Y^4}{\sin^2\theta}+{\sigma_Y^2}}}}d\theta.
	 \end{eqnarray}
\end{small}
\vspace{-3mm}

\noindent We upper bound (\ref{eqn:Pe})  by setting $\theta\!=\!\pi/2$ as  (see Appendix \ref{app:AppendixCC})

\vspace{-3mm}
\begin{small}
	\begin{eqnarray}\label{eqn:Peupper}
	\bar{P}^{ub}_e&=&\frac{\alpha{\psi}\mathrm{exp}(-\mu_Y^2/2\sigma_Y^2)}{2\sigma_Y}
	\frac{\mathrm{exp}\left(\mu_Y^2\big/\left({2\beta\bar{\gamma}\sigma_Y^4}+2\sigma_Y^2\right)\right)}{\sqrt{{\beta\bar{\gamma}}+{1}/{\sigma_Y^2}}}\nonumber\\
	&&\quad \quad \quad \times \Q{{\mu_Y}\big/{\sqrt{{\beta\bar{\gamma}\sigma_Y^4}+{\sigma_Y^2}}}}.
	\end{eqnarray} 
\end{small}
\vspace{-4mm}

\noindent\textit{\textbf{Remark 2:}}   The exact average   SER for $M=1$   can be derived by evaluating $\bar{P}_e=\E{\alpha\Q{\sqrt{\beta\Gamma^*}}}$ using $\cdf[\Gamma^*]{z}$ and invoking \cite[Eqn. (6.614.5)]{Gradshteyn2007} as follows: 

\vspace{-5mm}
\begin{small}
	\begin{eqnarray}\label{eqn:exact_SER}
	\bar{P}_e=\frac{\alpha}{2}-\frac{\alpha\,\delta}{2}\mathrm{exp}{\left(\delta\right)}\left(K_1(\delta)-K_0(\delta)\right),
	\end{eqnarray} 
\end{small}
\vspace{-4mm}

\noindent where $\delta=(4\beta\bar{\gamma}\lambda^2_1)^{-1}$.

\vspace{-2mm}
\subsection{Achievable diversity order}
The diversity order is  defined as the   negative slope of the outage probability or average SER versus the average SNR curve in a log-log scale as \cite{Wang2003a}

\vspace{-5mm}
\begin{small}
	 \begin{eqnarray}\label{eqn:Gd_def}
	 G_d = \lim_{\bar\gamma \rightarrow \infty} - \frac{\log  {P_{out}} }{\log{\gamma}} =  \lim_{\bar\gamma \rightarrow \infty} - \frac{\log  {\bar {P}_{e}} }{\log{\gamma}},
	 \end{eqnarray}
\end{small}
\vspace{-3mm}

\noindent from which useful information about how the outage probability or average SER decays in high SNR regime  can be obtained. Since the outage probability and the average SER  have identical diversity orders \cite{Wang2003a}, we first proceed our diversity order derivation by using  $P_{out}$.

In general,  the outage probability can be asymptotically approximated in the high SNR regime as 
$P_{out} \approx (O_c\bar \gamma)^{-G_d}$, where $G_d$ is the diversity order and  $O_c$ is a measure of the coding gain \cite{Wang2003a}. A single-polynomial approximation of $P_{out}$ in (\ref{eqn:P_out})  can be derived as (see Appendix \ref{app:AppendixDD}) 

\vspace{-5mm}
\begin{small}
	 \begin{eqnarray}\label{eqn:Poutasym}
	 P^\infty_{out}=\Omega_{op} \left(\frac{\gamma_{th}}{\bar{\gamma}}\right)^{G_d}+\mathcal{O}\left(\bar{\gamma}^{-(G_d+1)}\right),
	 \end{eqnarray}
\end{small}
\vspace{-3mm}

\noindent where the diversity order $G_d$ is    given by 

\vspace{-4mm}
\begin{small}
\begin{eqnarray}\label{eqn:diversity}
G_d=M,
\end{eqnarray}	 
\end{small}
\vspace{-6mm}

\noindent
where $M$ is the number of passive reflective elements in the IRS. In (\ref{eqn:Poutasym}), $\Omega_{op}=\xi\prod_{m=1}^M(\lambda^2_m(2M)!)^{-1}$, where the constant $\xi$ depends   on the coding/array gain. 

Similarly, an asymptotic approximation for the average SER in high SNR regime can be derived as (see Appendix \ref{app:AppendixDD})

\vspace{-4mm}
\begin{small}
	\begin{eqnarray}\label{eqn:Pe_asym}
	\bar{P}^{\infty}_e = \left(G_c \bar \gamma\right)^{-M}+\mathcal{O}\left(\bar{\gamma}^{-(G_d+1)}\right),
	\end{eqnarray} 
\end{small}
\vspace{-6mm}

\noindent where the coding gain is given by $ G_c=(\alpha\Omega_{op}2^{M-1}\Gamma(M+1/2)/(\sqrt{\pi}\beta^M))^{-1/M}$.
 
\textit{\textbf{Remark 3:}} The achievable diversity gain is in the order of the number of passive reflective embedded in the IRS even though both $S$ and $D$ are each equipped with a single-antenna. It is worth noting that each passive reflective element reconfigures  phases of incident waves such that  they add coherently at $D$.  
The direct SISO transmission between $S$ and $D$ permits only a unit diversity order. In conventional wireless systems, diversity gains can only be achieved by either transmit beamforming  or via receive combining   by employing multiple transmit/receive RF chains. However, the IRS is able to provide a significant diversity order by virtue of  just passive reflectors with reconfigurable phases.    

 \begin{figure}[!t]\centering \vspace{0mm}
 	\def\svgwidth{150pt} 
 	\fontsize{8}{4}\selectfont 
\begingroup%
  \makeatletter%
  \providecommand\color[2][]{%
    \errmessage{(Inkscape) Color is used for the text in Inkscape, but the package 'color.sty' is not loaded}%
    \renewcommand\color[2][]{}%
  }%
  \providecommand\transparent[1]{%
    \errmessage{(Inkscape) Transparency is used (non-zero) for the text in Inkscape, but the package 'transparent.sty' is not loaded}%
    \renewcommand\transparent[1]{}%
  }%
  \providecommand\rotatebox[2]{#2}%
  \newcommand*\fsize{\dimexpr\f@size pt\relax}%
  \newcommand*\lineheight[1]{\fontsize{\fsize}{#1\fsize}\selectfont}%
  \ifx\svgwidth\undefined%
    \setlength{\unitlength}{408.49222499bp}%
    \ifx\svgscale\undefined%
      \relax%
    \else%
      \setlength{\unitlength}{\unitlength * \real{\svgscale}}%
    \fi%
  \else%
    \setlength{\unitlength}{\svgwidth}%
  \fi%
  \global\let\svgwidth\undefined%
  \global\let\svgscale\undefined%
  \makeatother%
  \begin{picture}(1,0.34163772)%
    \lineheight{1}%
    \setlength\tabcolsep{0pt}%
    \put(0,0){\includegraphics[width=\unitlength,page=1]{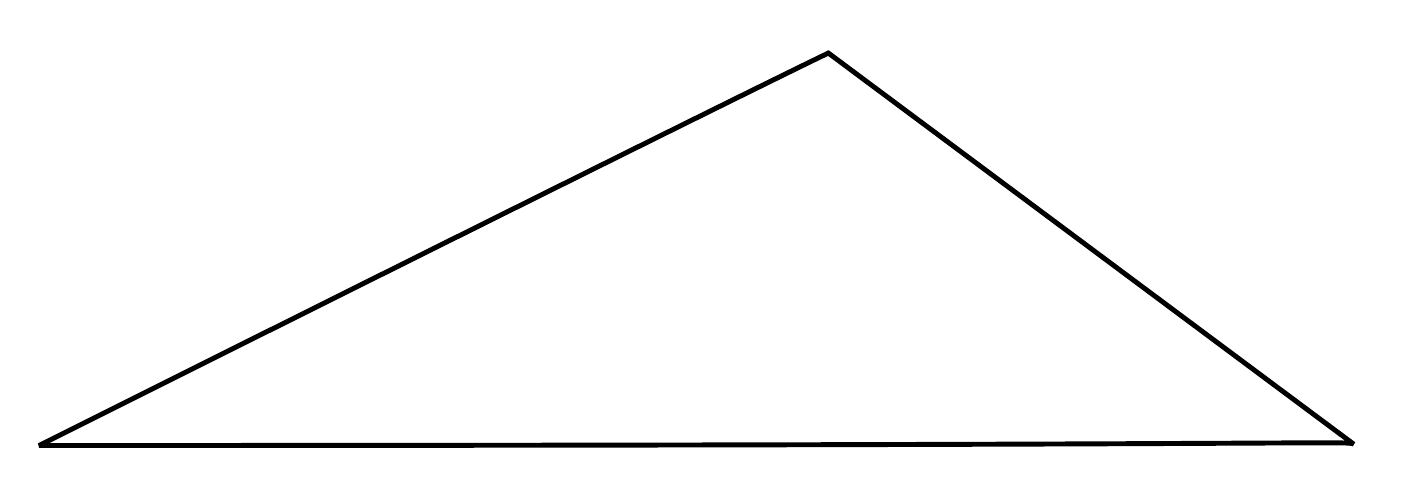}}%
    \put(0.45453325,0.04635897){\color[rgb]{0,0,0}\makebox(0,0)[lt]{\lineheight{1.25}\smash{\begin{tabular}[t]{l}$d_{SD}$\end{tabular}}}}%
    \put(0.7304572,0.21413722){\color[rgb]{0,0,0}\rotatebox{-33.592933}{\makebox(0,0)[lt]{\lineheight{1.25}\smash{\begin{tabular}[t]{l}$d_{DI}$\end{tabular}}}}}%
    \put(0.29421077,0.18328397){\color[rgb]{0,0,0}\rotatebox{25.893542}{\makebox(0,0)[lt]{\lineheight{1.25}\smash{\begin{tabular}[t]{l}$d_{SI}$\end{tabular}}}}}%
    \put(0,0){\includegraphics[width=\unitlength,page=2]{System_Model_Distances.pdf}}%
    \put(0.54610988,0.21251585){\color[rgb]{0,0,0}\makebox(0,0)[lt]{\lineheight{1.25}\smash{\begin{tabular}[t]{l}$\text{IRS}$\end{tabular}}}}%
    \put(0,0){\includegraphics[width=\unitlength,page=3]{System_Model_Distances.pdf}}%
    \put(0.0219988,0.0802977){\color[rgb]{0,0,0}\makebox(0,0)[lt]{\lineheight{1.25}\smash{\begin{tabular}[t]{l}$S$\end{tabular}}}}%
    \put(0.92902595,0.07744801){\color[rgb]{0,0,0}\makebox(0,0)[lt]{\lineheight{1.25}\smash{\begin{tabular}[t]{l}$D$\end{tabular}}}}%
  \end{picture}%
\endgroup%
 \vspace{-2mm}
 	\caption{Geometric placement of  nodes for numerical results.}\vspace{-5mm} \label{fig:system_model_sim}
 \end{figure}

\begin{figure}[!t]\centering\vspace{0mm}
	\includegraphics[width=0.4\textwidth]{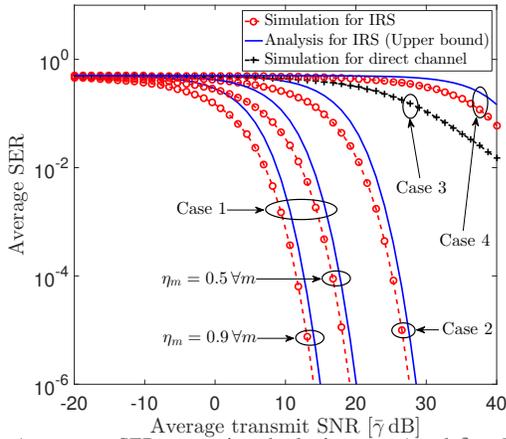}\vspace{-5mm}
	\caption{An average SER comparison by letting $\alpha=1$ and $\beta=2$ in (\ref{eqn:Pe}) for $M=32$. For all cases, $d_{SD}=100$\,m.  The distances for Case-1 to Case-4 are set to  $d_{SI}=d_{DI}=51$\,m, $d_{SI}=d_{DI}=80$\,m, and $d_{SI}=d_{DI}=170$\,m.    }
	\label{fig:BER_Comparison}\vspace{-5mm}
\end{figure}

 \begin{figure}[!t]\centering\vspace{-0mm}
 	\includegraphics[width=0.4\textwidth]{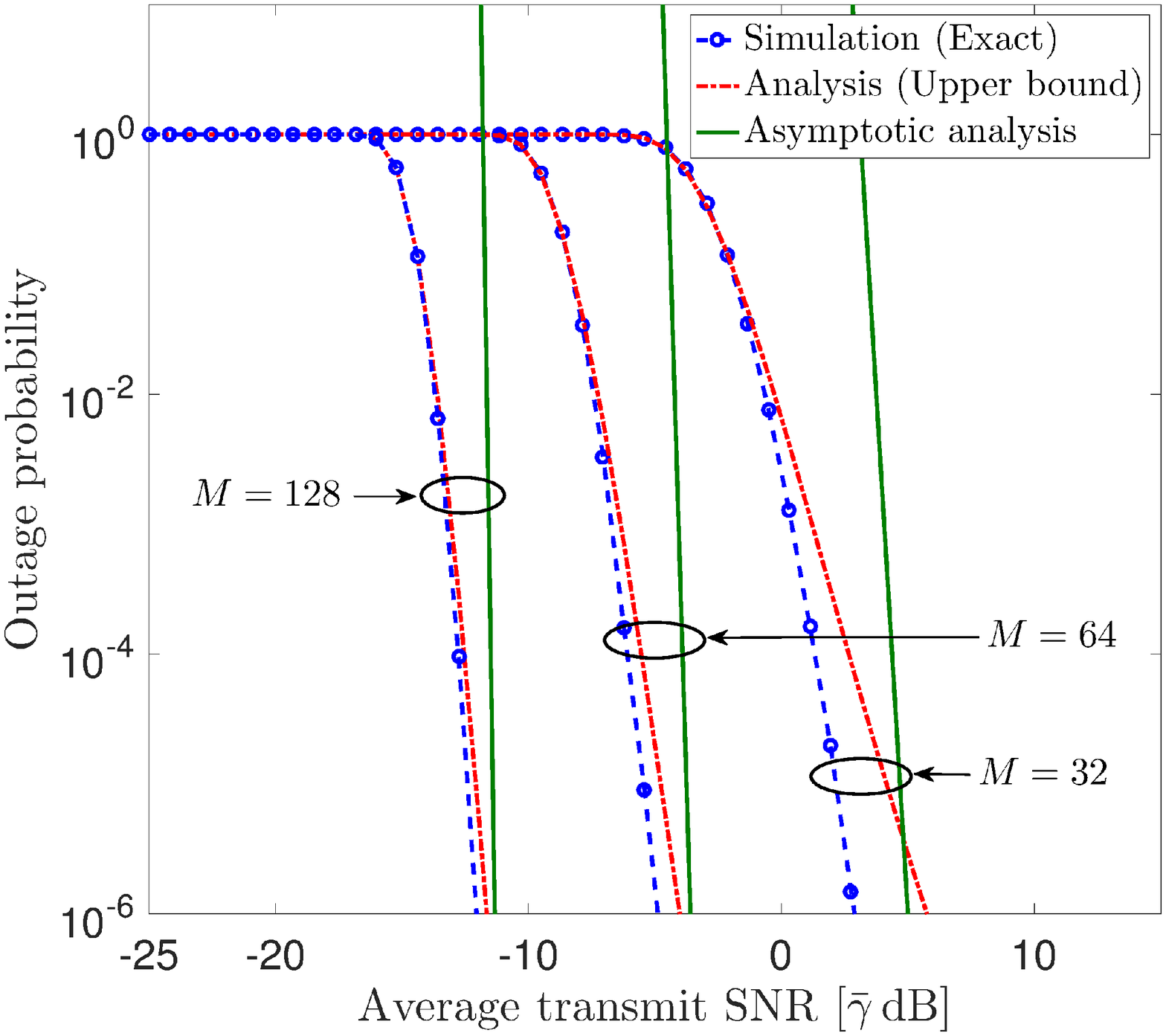}\vspace{-3mm}
 	\caption{The outage probability with $\gamma_{th}\!=\!0$\,dB, $d_{SI}\!=\!30$\,m and $d_{DI}\!=\!20$\,m. }
 	\label{fig:Fig_outage_probability}\vspace{-5mm}
 \end{figure}
 
\begin{figure}[!t]\centering\vspace{0mm}
	\includegraphics[width=0.4\textwidth]{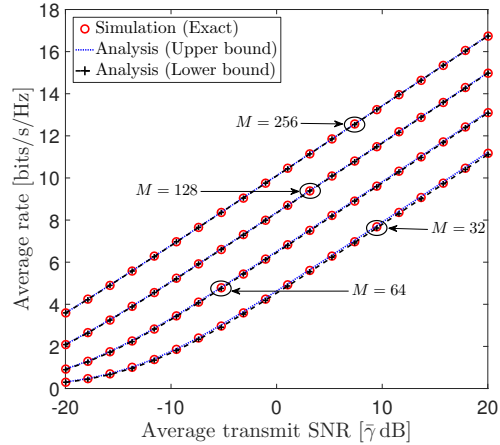}\vspace{-3mm}
	\caption{The average  rate with  $d_{SI}=30$\,m and $d_{DI}=20$\,m. }
	\label{fig:AchievableRate_vs_SNR}\vspace{-5mm}
\end{figure}

  \vspace{-2mm}
 \section{Numerical results}\label{sec:Numerical_resilts}
 
   \vspace{-2mm}
In simulations, the path-loss is modeled as $\zeta\,[\text{dB}]=\zeta_0+10\upsilon \log{d}$, where  $\zeta_0=42$\,dB is a reference path-loss,  $\upsilon=3.5$ is  the path-loss exponent and $d$ is the distance. The geometric placement of nodes is depicted in Fig. \ref{fig:system_model_sim}.  We set $\eta_m=\eta, \;\forall m$ for the sake of exposition.


{ Fig. \ref{fig:BER_Comparison} provides a comparison of the average SER  for the IRS-assisted communication set-up (Case-1 to Case-3)   with respect to a baseline    direct transmission between  $S$ and $D$ (Case-4). To highlight the performance gain of IRS over the direct transmission, in Case-1,  the IRS is placed such that $d_{SI}+d_{DI}\approx d_{SD}$. 
	Fig. \ref{fig:BER_Comparison} shows clearly that IRS-assisted communication outperforms its direct transmission counterpart.{ For example, at an  average SER of $10^{-1}$, IRS-assisted system provides an approximately 28.56\,dB (23.37\,dB) reduction in the average transmit SNR for $\eta_m=0.9\;(\eta_m=0.5)\;\forall m$ in comparison to the direct transmission (see Case-1 and Case-4).} When IRS is placed such that $d_{SD}<d_{SI}+d_{ID}<2d_{SD}$ (Case-2), the IRS-assisted communication can still provide a  considerable average  SER improvement compared to the direct transmission (see Case-2 and Case-4).
	 This performance gain is obtained by means of  the diversity gain rendered by smartly controlling the phase-shifts at each reflective element of the IRS to enable constructive addition of signals at $D$. However, when $d_{SI}+d_{ID}\gg d_{SD}$ (Case-3), the performance of IRS-assisted system is severely hindered  mainly due to the substantial increase in distance-dependent path-loss effects (see Case-3 and Case-4). }   	Fig. \ref{fig:BER_Comparison} also reveals that the average SER heavily depends on the reflection coefficient ($\eta_m$) of IRS elements. For instance, at an average SER of $10^{-4}$, an IRS with $\eta_{m}=0.9$ achieves a transmit SNR gain of {5.1}\,dB over that of an  IRS with $\eta_m = 0.5$.

{In Fig. \ref{fig:Fig_outage_probability}, the outage probability is plotted as a function of the average transmit SNR. Specifically,  the proposed outage upper bound in (\ref{eqn:Peupper}) and its high SNR counterpart in (\ref{eqn:Poutasym}) are plotted  together with the exact Monte-Carlo simulations. Fig. \ref{fig:Fig_outage_probability} clearly depicts that the tightness of our proposed analytical outage  improves in large $M$ regime. Counter-intuitively, it becomes looser in the  high SNR regime. Nevertheless, our high SNR outage approximation tends to be asymptotically exact, and hence, it can be used to analytically quantify the diversity order and the corresponding outage performance in high SNR regime. Thus, the asymptotic outage curves in Fig. \ref{fig:Fig_outage_probability} verify that the IRS-assisted set-up achieves an $M$th order diversity gain.     Fig. \ref{fig:Fig_outage_probability} reveals that the outage probability can be lowered  significantly by increasing $M$.  For instance,  a quadruple and double increments in $M$   provide 13\,dB and 7\,dB reductions in the average transmit SNR to attain the same  outage probability of  $10^{-4}$.
}

In Fig. \ref{fig:AchievableRate_vs_SNR}, the average achievable rate is plotted for different  number of reflective elements  at IRS as  $M=[32, 64, 128, 256]$. The exact achievable rate is plotted from (\ref{eqn:Optimal_SNR_Rate}) by using Monte-Carlo simulations. The analytical curves for the upper and lower bounds are plotted by using (\ref{eqn:Optimal_SNR_Rate_ub}) and (\ref{eqn:lowerbound_ana}), respectively. Fig. \ref{fig:AchievableRate_vs_SNR} clearly illustrates that our upper bound is tight even for a moderately large number of IRS elements such as $M=32$. Moreover, the tightness of our lower bound improves with increasing $M$.  Both upper and lower bounds approach  the exact simulation  when the number of IRS elements grows large (see $M=256$ case).

\vspace{-2mm}
\section{Conclusion}\label{sec:conclusion}
The performance of IRS for wireless communication has been  investigated by deriving the outage probability, average SER and achievable rate bounds and approximations. The accuracy of these metrics becomes tighter when the number of reflective elements in the IRS grows large. By deriving single-polynomial high SNR approximations of the CDF and PDF of the SNR, the achievable diversity order has been quantified. This high SNR analysis reveals that the achievable diversity order is equal to the number of passive reflective elements. A set of rigorous numerical results is provided to validate our performance analysis and to obtain useful insights about employing  IRS for boosting the performance of next-generation  wireless systems.


\vspace{-2mm}
\appendices

\section{Derivation of the CDF of $\gamma$ in (\ref{eqn:Optimal_SNR_CDF})} \label{app:AppendixAA} 
We rewrite   $\gamma\!=\!\bar{\gamma}Y^2$, where $Y\!=\!\sum_{m=1}^M X_m$ and  $X_m\!=\!\rho_m\nu_m\eta_m$. Here, $\rho_m$ and $ \nu_m$ in (\ref{eqn:SNR_Optimal}) are two independent Rayleigh distributed RVs  with parameters $\zeta_{h_m}/2$ and $\zeta_{g_m}/2$, respectively. The $k$th moment of  $X_m$ is given by  \cite{Salo2006}

\vspace{-4mm}
\begin{small}
\begin{eqnarray}
\bar{x}_{k,m}=\E{X^k_m}=(\zeta_{h_m}\zeta_{g_m}\eta^2_m)^{k/2}\left[\GammaFn{k/2+1}\right]^2.
\end{eqnarray}
\end{small}
\vspace{-4mm}

\noindent By invoking CLT, the PDF of $Y$ can be approximated by a Gaussian distribution in moderately large   $M$ regime;  $Y\sim\mathcal N(\mu_Y,\sigma_Y^2)$,   where 
$\mu_Y$ and $\sigma_Y^2$ are given by

\vspace{-4mm}
\begin{small}
\begin{eqnarray}
\!\!\!\!\!\!\!\mu_Y&=&\sum_{m=1}^M\bar{x}_{1,m} \quad\text{and}\quad 
\;\;\;
\sigma_Y^2=\sum_{m=1}^M\left(\bar{x}_{2,m}-(\bar{x}_{1,m})^2\right).
\end{eqnarray}
\end{small}
\vspace{-4mm}

\noindent 
The final expressions for  $\mu_Y$ and $\sigma_Y^2$ are given in (\ref{eqn:mean}). Thereby, the PDF of $Y$ can be written as

\vspace{-4mm}
\begin{small}
\begin{eqnarray}\label{eqn:fy}
f_{Y}(y)= 
{\psi}\mathrm{exp}\left(-{(y-\mu_Y)^2}\big/{2\sigma_Y^2}\right)\big/{\sqrt{2\pi\sigma_Y^2}} ,\;\;y\geq 0, 
\end{eqnarray} 
\end{small}
\vspace{-4mm}

\noindent  where  $f_{Y}(y)=0$ for $y<0$,  $\psi$ is a normalization coefficient, which  is defined in (\ref{eqn:varaiance}), and computed using the fact that  $\int_{-\infty}^{\infty}f_{Y}(u)du=1$. The CDF of $Y$ is derived as 

\vspace{-4mm}
\begin{small}
\begin{eqnarray}\label{eqn:Fy}
F_{Y}(y)&=&\int_{-\infty}^{y}f_{Y}(u)du
=1-\psi\Q{(y-\mu_Y)/\sigma_Y},
\end{eqnarray}
\end{small}
\vspace{-4mm}

\noindent  for $y\geq 0$ and $F_{Y}(y) = 0$ elsewhere. The CDF of $\gamma=\bar{\gamma}Y^2$ can be derived via transformation of RVs   as 

\vspace{-4mm}
\begin{small}
\begin{eqnarray}\label{eqn:Fgamma}
F_{\gamma}(z)&=&\mathrm{Pr}\left(\gamma\leq z\right)=
\mathrm{Pr}\left(-\sqrt{z/\bar{\gamma}}\leq Y\leq \sqrt{z/\bar{\gamma}}\right)\nonumber\\
&=&F_{Y}(\sqrt{z/\bar{\gamma}})-F_{Y}(-\sqrt{z/\bar{\gamma}}) \;\; \text{for } z\geq0,
\end{eqnarray}
\end{small}
\vspace{-4mm}

\noindent  and $F_{\gamma}(z) = 0$ otherwise. By substituting (\ref{eqn:Fy}) to (\ref{eqn:Fgamma}), the CDF for $\gamma$ can be derived as given in (\ref{eqn:Optimal_SNR_CDF}). 
  \setcounter{mycounter}{\value{equation}}
  \begin{figure*}[!t] 
  	\addtocounter{equation}{11}
  	\begin{small}
  	\begin{eqnarray}\label{eqn:mgf}
\mathcal M_{X_m}(s)&=&\E{\exp{-sX_m}}=\left({{(s\lambda_m)}^2-1}\right)^{-1}\left[{s\lambda_m}\ln\left({s\lambda_m+\sqrt{\left({s}\lambda_m\right)^2-1}}\right)\left({\sqrt{(s\lambda_m)^2-1}}\right)^{-1/2}-1\right]
  	\end{eqnarray}
  \end{small}
  
  	\vspace{-6mm}
  	\hrulefill
  	
  	\vspace{-6mm}
  \end{figure*}
  \setcounter{equation}{\value{mycounter}} 

\vspace{-2mm}
\section{Derivation of $\mathcal R_{ub}$ in (\ref{eqn:Optimal_SNR_Rate_ub}) and $\mathcal R_{lb}$ in (\ref{eqn:lowerbound_ana}) } \label{app:AppendixBB} 

By using $\gamma=\bar{\gamma}Y^2$, (\ref{eqn:Optimal_SNR_Rate}) can be rewritten as  $\mathcal R=\E[Y]{\log[2]{1+\bar{\gamma}Y^2}}$. By noticing that  $\log[2]{1+x^2}$ is a concave function for $x>0$, we invoke the Jensen's inequality  \cite{Proakis2007} to derive an upper bound for  $\mathcal R$  as 

\vspace{-4mm}
\begin{small}
\begin{eqnarray}\label{eqn:upper}
\mathcal R\leq \mathcal R_{ub}= \log[2]{1+\bar{\gamma}\E{Y^2}}. 
\end{eqnarray}
\end{small}
\vspace{-4mm}

\noindent where $\E{Y^2}$ can be  derived as $\E{Y^2}=\mu_Y^2+\sigma_Y^2$. By substituting $\E{Y^2}$ into (\ref{eqn:upper}), the desired upper bound for the average achievable rate can be derived as (\ref{eqn:Optimal_SNR_Rate_ub}).

Next, the  derivation of $\mathcal R_{lb}$ in (\ref{eqn:lowerbound_ana}) is outlined. To begin with, by applying the Taylor series expansion of $1/\gamma$ around $\E{\gamma}$ \cite{Gradshteyn2007},   the term $\E{1/\gamma}$ in (\ref{eqn:lwerbound})  can be approximated as \cite{Zhang2014}

\vspace{-4mm}
\begin{small}
\begin{eqnarray}\label{eqn:approx_gamma}
\E{{1}\Big/{\gamma}}\approx{1}\Big/{\E{\gamma}}+{\Var{\gamma}}\Big/{\left[\E{\gamma}\right]^3}.
\end{eqnarray}
\end{small}
\vspace{-4mm}

\noindent Since $\gamma$ follows a non-central chi-squared distribution with one degree-of-freedom,   mean and variance are given by  \cite{Papoulis2002}

\vspace{-4mm}
\begin{small}
	 \begin{subequations}
	 	\begin{eqnarray}
	 	&&\!\!\!\!\!\!\!\!\!\!\!\!\!\!\!\!\!
	 	\E{\gamma}=\bar{\gamma}(\sigma_Y^2+\mu_Y^2)= \left(\bar{\gamma}\sum_{m=1}^M\lambda^2_{m}\right)\left[\frac{(16-\pi^2)(1+\kappa)}{4}\right]\!\!,\label{eqn:meanchi}\\
	 	&&\!\!\!\!\!\!\!\!\!\!\!\!\!\!\!\!\!
	 	\Var{\gamma}=2\sigma_Y^2\bar{\gamma}^2(\sigma_Y^2\!+\!2\mu_Y^2)\nonumber\\
	 	&&=
	 	\left(\bar{\gamma}\sum_{m=1}^M\lambda^2_{m}\right)^2\left[\frac{(16-\pi^2)^2(1+2\kappa^2)}{8}\right].
	 	\label{eqn:varchi}
	 	\end{eqnarray}
	 \end{subequations}
\end{small}
\vspace{-4mm}

\noindent By first replacing $\E{\gamma}$ and $\Var{\gamma}$   in (\ref{eqn:approx_gamma}) via   (\ref{eqn:meanchi}) and (\ref{eqn:varchi}), respectively, and then by substituting the resultant expression into (\ref{eqn:lwerbound}), $\mathcal R_{lb}$ can be approximated as shown in (\ref{eqn:lowerbound_ana}).

\section{Derivation of $\bar{P}_e$ in (\ref{eqn:Pe})}\label{app:AppendixCC} 

By substituting $\gamma\!=\!\bar{\gamma}Y^2$ into $\bar{P}_e \!\approx\! \E{\alpha\Q{\sqrt{\beta\gamma}}}$, we have 

\vspace{-4mm}
\begin{small}
	 \begin{eqnarray}\label{eqn:P_e}
	 \bar{P}_e=\int_{0}^{\infty}\alpha \Q{x\sqrt{\beta \bar{\gamma}} }f_{Y}(x)dx,
	 \end{eqnarray} 
\end{small}
\vspace{-3mm}

\noindent
where $f_Y(y)$ is given in (\ref{eqn:fy}). By substituting $f_Y(y)$  and after  several mathematical manipulations, (\ref{eqn:P_e}) can be simplified as 

\vspace{-3mm}
\begin{small}
	 \begin{eqnarray}\label{eqn:Pe2}
	 \bar{P}_e=\Delta\int_{0}^{\infty}\Q{\sqrt{a}x}\mathrm{exp}\left(-(bx^2-2cx)\right)dx,
	 \end{eqnarray}
\end{small}
\vspace{-4mm}

\noindent
where $\Delta$, $a, b$, and $c$ coefficients are given by

\vspace{-4mm}
\begin{small}
\begin{eqnarray}
&&\Delta=\alpha{\psi}\mathrm{exp}(-\mu_Y^2/2\sigma_Y^2)/{\sqrt{2\pi\sigma_Y^2}}, \;\; a=\beta\bar{\gamma},\label{eqn:para1}\\
&&b=1/2\sigma_Y^2\;\; \text{and} \;\; c=\mu_Y/{2\sigma_Y^2}.\label{eqn:para2}
\end{eqnarray}
\end{small}
\vspace{-4mm}

\noindent Here,  $\Q{\sqrt{a}x}$, in (\ref{eqn:Pe2}) can be alternatively written as \cite{Gradshteyn2007}

\vspace{-3mm}
\begin{small}
	 \begin{eqnarray}\label{eqn:alter_Q}
	 \Q{\sqrt{a}x}=\frac{1}{\pi}\int_{0}^{\pi/2}\mathrm{exp}\left(-\frac{ax^2}{2\sin^2\theta}\right)d\theta.
	 \end{eqnarray}
\end{small}
\vspace{-4mm}

\noindent By substituting (\ref{eqn:alter_Q}) into (\ref{eqn:Pe2})  and by applying several mathematical manipulations, we have 

\vspace{-3mm}
\begin{small}
	 \begin{eqnarray}\label{eqn:doubleint}
	 \!\!\!\!\!\!\!\!\!\bar{P}_e\!=\!\frac{\Delta}{\pi}\!\!\int_{0}^{\pi/2}\!\!\!\!
	 \int_{0}^{\infty}\!\!\!\mathrm{exp}\left(\!\!-\!\left[\left(\frac{a}{2\sin^2\theta}\!+\!b\right)\!x^2\!+\!2cx\right]\right)dxd\theta.
	 \end{eqnarray}
\end{small}
\vspace{-4mm}

\noindent
By invoking \cite[2.33.1]{Gradshteyn2007}, the inner integral of (\ref{eqn:doubleint}) can be evaluated, and then, (\ref{eqn:doubleint})  reduces to 

\vspace{-3mm}
\begin{small}
\begin{eqnarray}
\!\!\!\!\!\!\!\!\!\bar{P}_e\!&=&\!\frac{\Delta}{\sqrt{\pi}}\int_{0}^{\frac\pi 2}\!\frac{\mathrm{exp}\left(c^2\!\big/\!\!\left(\frac{a}{2\sin^2\theta}\!+\!b\right)\right)}{\sqrt{\frac{a}{2\sin^2\theta}+b}}\Q{\!\!-\frac{\sqrt{2}c}{\sqrt{\frac{a}{2\sin^2\theta}+b}}\!}d\theta. 
\end{eqnarray}
\end{small}
\vspace{-3mm}

\noindent By substituting $\Delta$, $a$, $b$, and $c$ from (\ref{eqn:para1})-(\ref{eqn:para2}),   $\bar{P}_e$ can be written as (\ref{eqn:Pe}).

\section{Derivation of diversity order in (\ref{eqn:Poutasym})}\label{app:AppendixDD} 
The PDF of a product of two independent  Rayleigh RVs,  $X_m=\rho_m\nu_m\eta_m$, is given by \cite{Salo2006}

\vspace{-3mm}
\begin{small}
\begin{eqnarray}\label{eqn:PDFXm}
f_{X_m}(x)=x/\lambda^2_mK_{0}\left(x/\lambda_m\right),\quad x\geq 0,
\end{eqnarray}
\end{small}
\vspace{-5mm}

\noindent
and $f_{X_m}(x) = 0$ for $x<0$. The moment generating function (MGF) of  $X_m$ can be derived as (\ref{eqn:mgf}) at the top of this page  \cite[6.624.1]{Gradshteyn2007}. Since  $\{X_1,\cdots,X_m\cdots, X_M\}$ are independent RVs, the MGF of $Y=\sum_{m=1}^M X_m$ can be derived as $\mathcal M_{Y}(s)=\prod_{m=1}^{M}\mathcal M_{X_m}(s)$ \cite{Papoulis2002}.

The order of smoothness of the PDF of $\gamma = \bar \gamma Y^2$ at the origin can be used to 
investigate the asymptotic behavior of the outage probability or average SER at the high SNR regime \cite{Wang2003a}. The corresponding  order of smoothness of $f_{\gamma}(x)$  at the origin can be translated into the decaying order of the pertinent MGF, $M_{\gamma}(s)$,  which decays as a function of $s$ \cite{Wang2003a}.
To this end,   $\mathcal M_{Y}(s)$ in (\ref{eqn:mgf}) can be approximated when   $s\rightarrow \infty$ as
\addtocounter{equation}{1} 

\vspace{-4mm}
\begin{small}
	 \begin{eqnarray}\label{eqn:approx_infty}
	 \!\!\!\!\!\!\lim_{s \rightarrow \infty}\mathcal M_{Y}(s)\!=\!\mathcal M^\infty_{Y}(s)\!\approx\!\prod_{m=1}^M\!\!\frac{1}{(\lambda_m s)^2}\!\left[\xi(\lambda_m s)^{1/\xi}\!-\!\xi'\right],
	 \end{eqnarray}
\end{small}
\vspace{-3mm}

\noindent
where $\xi$ is a large number such that $\ln{\left({\lambda_m s}\right)}\approx \xi(\lambda_m s)^{1/\xi}-\xi$ and  $\xi'=\ln2-\xi$. The approximation in (\ref{eqn:approx_infty}) becomes tight for large $\xi$ values. Hence, for large values of $\xi$, we have 

\vspace{-3mm}
\begin{small}
\begin{eqnarray}\label{eqn:asumpMGFy}
\mathcal M^\infty_{Y}(s)= \Omega_Y/s^{2M}+\mathcal O(s^{-{2M}-\epsilon}),\;\; \epsilon>0,
\end{eqnarray}
\end{small}
\vspace{-4mm} 

\noindent 
where $\Omega_Y=\xi\prod_{m=1}^M(\lambda_m^2)^{-1}$. By invoking inverse Laplace transform \cite{Gradshteyn2007} on (\ref{eqn:asumpMGFy}), the PDF of $Y$ can be approximated by a single polynomial term for $y\rightarrow 0^+$ (i.e., $y$ approaches  zero from above) as 

\vspace{-4mm}
\begin{small}
\begin{eqnarray}\label{eqn:pdfY}
f^{ 0^+}_{Y}(y)=\Omega_Y/(2M!) y^{2M-1}+\mathcal O(y^{2M-1+\epsilon}), 
\end{eqnarray}
\end{small}
\vspace{-4mm} 

\noindent 
for $\epsilon>0$. From (\ref{eqn:pdfY}),  the corresponding  CDF can be readily derived  as $F^{0^+}_Y(y)=\int_{0}^{y}f^{0^+}_Y(t)dt$. Then, by performing  the variable transformation, $y=\sqrt{z/\bar{\gamma}}$, a single polynomial  approximation of the CDF of $\gamma$ can be derived as

\vspace{-3mm}
\begin{small} 
\begin{eqnarray}\label{eqn:Fgammaasymp}
\!\!\!\!\!\!\!\!\!
F^{ 0^+}_{\gamma}(z)=\Omega_{op} \left({z}/{\bar{\gamma}}\right)^{M}\!\!+\!\mathcal O\left({\left({z}/{\bar{\gamma}}\right)}^{M+1+\epsilon}\right), \;\; \text{for} \;\;\epsilon>0.
\end{eqnarray}
\end{small}
\vspace{-4mm} 

\noindent 
where $\Omega_{op}=\Omega_Y/(2M!)$. 
Then, the asymptotic outage probability can be derived as $P^{\infty}_{out}=F^{ 0^+}_{\gamma}(\gamma_{th})$ as in (\ref{eqn:Poutasym}).

{Next, the derivation of asymptotic average  SER (\ref{eqn:Pe_asym}) is outlined.  An integral for computing the  average SER is given by $\bar{P}_e=\alpha\sqrt{\beta}/(2\sqrt{2\pi})\int^{\infty}_{0}x^{-1/2}\mathrm{exp}(-\beta x/2)F_{\gamma}(x)dx$ \cite{Amarasuriya2010b}. By substituting (\ref{eqn:Fgammaasymp}) into $\bar{P}_e$, an asymptotic approximation for the average SER can be derived as

\vspace{-3mm}
\begin{small}  
	\begin{eqnarray}\label{eqn:asympPe}
	\bar{P}^{\infty}_e=\frac{\alpha\Omega_{op}}{2\bar{\gamma}^M}\sqrt{\frac{\beta}{2\pi}}\int_{0}^\infty x^{M-\frac{1}{2}}\mathrm{exp}(-\beta x/2) dx.
	\end{eqnarray}
\end{small}
\vspace{-4mm} 

\noindent 
	By substituting $t=\beta x/2$ into (\ref{eqn:asympPe}), and evaluating the integral via \cite[Eqn. (8.310.1)]{Gradshteyn2007}, the asymptotic  average SER at high SNR regime can be derived as (\ref{eqn:Pe_asym}). }

  \linespread{1.0}

\end{document}